# Leveraging Large Language Models for Actionable Course Evaluation Student Feedback to Lecturers

**M Zhang**
Aalborg University
Copenhagen, Denmark
https://orcid.org/0000-0003-1218-5201

**E D Lindsay** [1]
Aalborg University
Aalborg, Denmark
https://orcid.org/0000-0003-3266-164X

**F B Thorbensen**
Aalborg University
Copenhagen, Denmark
ORCID

**D B Poulsen**
Aalborg University
Copenhagen, Denmark
https://orcid.org/0000-0001-9623-0748

**J Bjerva**
Aalborg University
Copenhagen, Denmark
https://orcid.org/0000-0002-9512-0739

## ABSTRACT

End of semester student evaluations of teaching are the dominant mechanism for providing feedback to academics on their teaching practice. For large classes, however, the volume of feedback makes these tools impractical for this purpose. This paper explores the use of open-source generative AI to synthesise factual, actionable and appropriate summaries of student feedback from these survey responses. In our setup, we have 742 student responses ranging over 75 courses in a Computer Science department. For each course, we synthesise a summary of the course evaluations and actionable items for the instructor. Our results reveal a promising avenue for enhancing teaching practices in the classroom setting. Our contribution lies in demonstrating the feasibility of using generative AI to produce insightful feedback for teachers, thus providing a cost-effective means to support educators' development. Overall, our work highlights the possibility of using generative AI to produce factual, actionable, and appropriate feedback for teachers in the classroom setting.

---

[1] *E D Lindsay, edl@plan.aau.dk*

1. INTRODUCTION

Feedback is a powerful means of supporting learning (Hattie 2008) to inform the learner about their actual state of performance (Narciss 2008). Academics spend significant energy providing students with quality feedback on their work; but the pathways in which academics themselves receive feedback on their teaching are less supported. There are a range of approaches, such as structured university pedagogy programs and peer reviews from colleagues, but the dominant form of feedback to academics is the end of semester student evaluation.

Student teaching evaluations usually comprise a series of quantitative Likert-scale items as well as open-ended survey questions. The intention is that every student has the opportunity to contribute meaningful feedback about their learning experience, but as class sizes grow larger this feedback becomes less and less workable. For large classes the volume of open-ended feedback becomes unworkable, leading to a reliance upon the averages of the quantitative questions. These in turn become useful only if their values are very high (for the purpose of recognising teaching excellence) or very low (for the purpose of identifying needed interventions).

Recent breakthroughs in natural language processing (NLP), and in particular large language models (LLMs) have shown to have influenced a wide variety of research fields (Wahle et al. 2023), including education (Kasneci et al. 2023). These models show a remarkable ability to synthesise large quantities of text, taking as input long text sequences and generate text depending on the likelihood of the next token in the sequence. One can also tune these models to be able to take instructions and generate text based on these instructions.

There is a wide array of previous work on automated feedback, in the context of student learning, peer-to-peer learning, or peer reviewing. To elaborate, they are focused on comprehensive overviews of peer-to-peer feedback (Bauer et al. 2023), using sentence similarity methodologies for aligning answers to open responses in mathematical questions (Botelho et al. 2023), using language models to assist students improve their critical thinking skills by argumentation (Guerraoui et al. 2023), benchmarking whether GPT-style models (Brown et al. 2020; Achiam et al. 2023) align with feedback on research papers (Liang et al., 2023) and using LLMs in-the-loop during student programming assignments (Pankiewicz and Baker 2023).

In the context of automated feedback tools for teachers, previous studies have extensively explored automated feedback tools, which provide insights into student engagement and progress monitoring (Schwarz et al., 2018; Aslan et al., 2019; Alrajhi et al., 2021). These tools empower educators by offering analytics for intervention when necessary. Recent advancements in NLP enable feedback on educators' classroom discourse, promoting self-reflection and instructional enhancement in real time (Samei et al., 2014; Kelly et al., 2018; Jensen et al., 2020; Suresh et al. 2021; Demszky and Liu, 2023; Demszky et al. 2023). Wang and

Demszky (2023) investigate whether ChatGPT can be used for generating feedback that is useful for teacher development. However, to the best of our knowledge, previous work has not looked at using course evaluations as a feedback source. Another line of work is the usage of LLMs for coding student evaluations (Katz et al. 2024), the difference between this and our work is that we use the student evaluations directly as a source of critical information to give feedback on for the instructor of a course.

In this work, we explore whether such models can be used to synthesise meaningful feedback summaries from open-ended student evaluation responses. Therefore our key research question is:

*RQ: How can LLMs be applied to automatically synthesise student feedback in a manner that is factual, actionable, appropriate, and generates value for individual lecturers?*

In doing so, we must ensure that the output of our models demonstrates three key features, inspired by previous work (Wang and Demszky 2023; Wang et al. 2023; Guo et al. 2023; Chang et al. 2023):

1. **Factuality** – whether the model accurately generates feedback according to the input text, i.e., course evaluation, and does not hallucinate irrelevant information.
2. **Actionability** – that it is feedback that could be actioned by the lecturer, instead of giving a summarisation of the course evaluations.
3. **Appropriateness** – if the feedback that addresses teaching rather than being personal in nature. For example, if a student writes down an unwanted comment, this would not be echoed by the model.

This work was conducted at the Computer Science department of Aalborg University. Our dataset comprises student evaluations of 75 courses ranging across the five different year levels of the degree programs within the Department. Response rates varied from a single response in a course up to a maximum of 44 responses per course, with a total of 742 responses in the dataset. The dataset was drawn from a Covid-19-era teaching semester.

## 2. OUR MODEL

In this work, we used Llama2 (7B; Touvron et al. 2023) as the core LLM to synthesise the student feedback. Llama2 is an open-source 7-billion-parameter language model based on the Transformer architecture (Vaswani et al. 2017) and it generates text autoregressively.

We used a *zero-shot inference* (direct *prompting*) approach, relying on the ability of the model to make predictions or perform tasks in the absence of explicit training data pertaining to those specific tasks (Romera-Paredes and Torr 2015; Xian et al. 2017) and not showing any examples.

The decision not to fine-tune the model allows for a simpler and faster implementation process. It also allows for a transferable implementation, because it does not rely upon an extensive (and expensive) training or alignment process in order to achieve its results. In doing so, however, we forego the improved performance that tailoring the model to our context would provide.

We adopted a two-stage approach to synthesising the student feedback into an actionable summary. In the first stage we concatenated all of the student feedback for a particular course, and prompted the model to summarise that feedback:

*Prompt_1: "Summarise, to a maximum of {X} tokens this text that is based on course evaluations: {INPUT}"*

Where "X" number of tokens is determined by a heuristic so that the total input text length does not exceed the working capacity of the model. Having made a summary, we then prompted the model to provide actionable feedback for the lecturer:

*Prompt_2: "You are now an actionable feedback bot. Give actionable feedback, based upon these summarised course evaluations, to the instructor of the course. Leave out names that could identify entities. Make sure that the feedback is factual, actionable, and appropriate to the instructor: {SUMMARISATION}"*

This process resulted in 75 feedback summaries, one per unique course in our dataset.

## 3. RESULTS

In this section, we show several excerpts of the output of the LLM and consolidate emergent formatting standards of the LLM. The input of the LLM is a mixture of Danish and English course evaluations. We get the output of the LLM in free text format. We notice there were some issues with the translations between Danish and English, most likely due to our Llama2-based model, being an English-based model.

We observe that there are three key formats the model generates output given our input, without intentionally prompting for these formats. Overall, the majority of the emergent formats blended feedback with comments and listed the feedback points one by one (first format). A second format generated a direct response to the instructor, by listing the strengths, areas for improvement, positive feedback, and recommendations making it more verbose than the first format. In the third format, the model first presents a summary of the key aspects of the feedback (both positive and negative points) and gives the recommended actions in the second paragraph.

Decoupling the issues from the suggestions allows a holistic view of the class to emerge, and ensures that the focus of the feedback is on being actionable, rather than focussing on suggesting actions to the academic. This has similarities to the Chain of Thought reasoning abilities of LLMs (Wei et al. 2022; Zhang et al. 2022), where a series of intermediate reasoning steps improves the LLM on complex reasoning.

In Table 1, we show illustrative examples of the output of the model, specifically of the first and third format.

*Table 1. **Illustrative Excerpts.** We show examples of the generated output of the LLM. To save space, note that these outputs are excerpts of the full output of the LLM.*

| **Generated Output of the Model** |
|---|
| (1) Based on the course evaluations, here are actionable feedback points to improve the "flipped classroom" model: <br><br> 1. Workload Management: Students found the workload excessive, suggesting a reduction in assignments or more breaks. <br> 2. Adjusting Difficulty Levels: Some assignments were too challenging, suggesting a review and adjustment for consistency. <br> 3. Providing Additional Resources: Students suggested additional resources like past assignment reviews and solutions for better understanding. <br> 4. Peer Learning: Encourage interaction through group work or peer-to-peer activities. <br> 5. Exam Structure: Consider revising exams to align better with student needs, possibly integrating a separate programming course. <br> 6. Code Review: Incorporate more code review activities to demonstrate real-world implementation. <br><br> Overall, while the flipped classroom model was beneficial, addressing these points can enhance the course experience. |
| (3) Dear Instructor, <br><br> Here's a concise summary of the key feedback points derived from the course evaluations: <br><br> 1. Course Materials: The structured nature of the book and exercises received praise for aiding comprehension, with detailed examples enhancing understanding. <br> 2. Exam Preparation: The exam tasks practiced during the course were beneficial, though there's potential for improvement in actively listening to and incorporating student feedback. <br> 3. Communication Style: Some challenges arise from the lecturer's communication style, exacerbated by distance due to COVID restrictions. Utilizing slides for better visibility and adapting communication methods could address this. <br> 4. Instruction Clarity: Unclear or irrelevant instructions during lectures led to student confusion and dissatisfaction. <br> 5. Project Feedback: Students felt project feedback was overly critical, lacking empathy towards project challenges. <br><br><br> Actionable Feedback Suggestions: <br><br> 1. Enhance Communication Style: Work on improving communication to cater to all students, possibly through slower speech or visual aids. |

> 2. Clarify Instructions: Ensure instructions provided during lectures are clear and concise to prevent confusion.
> 3. Offer More Feedback: Encourage early project sharing for better understanding of student challenges and tailored feedback.
> 4. Utilize Online Resources: Integrate engaging online materials to supplement lectures and boost student engagement.
> 5. Adjust Seating Arrangements: Where possible, consider seating students closer together during COVID restrictions to mitigate visibility issues with slides.
>
> Addressing these suggestions can foster a more inclusive and effective learning environment for students.

## 4. EVALUATION OF THE OUTPUTS

A team of three human evaluators rated the model outputs on a five-point Likert scale, considering the three dimensions: Factuality, Actionability and Appropriateness. Overall, there was good agreement between the evaluators, with the majority of evaluations receiving the same or similar ratings on the Likert scale.

The process of reviewing divergent ratings revealed interpretation differences on the part of the evaluators, and the process of reconciling these differences identified key themes for the feasibility of this NLP approach.

### 4.1. Observations regarding Factuality.

Overall, the output of the model was largely agreed to be factually correct—it was indeed a synthesis of the input feedback from the students. The model was consistent in the amount of output it would generate, which presented challenges when dealing with the smallest and largest sets of input data.

For the smallest samples (there are seven courses with only one respondent) the output would reflect the input; but it would also include additional feedback not arising from the input. While these were largely benign in nature, they did not actually reflect factual feedback from the students.

These false positives appear to have mostly been introduced in the second step of the model. The first step summarisation of the inputs seems to be an accurate summary; it is the prompt to provide actionable feedback that triggers the model to generate additional comments not based upon the input.

For the largest samples the model suffered from not having pedagogical expertise, and as such being unable to correctly prioritise which input comments most needed to propagate to the output. This challenge was most pronounced where there were contradictory comments in the dataset. The model could identify that students were positive or negative towards an aspect of the teaching, but it struggled to convey the inconsistency among students, instead often only reporting one of the two sentiments.

## 4.2. Observations regarding Actionability.

The model was prompted to provide actionable feedback to the instructor, and for the most part the output provided suggestions to the academic based on the summaries. Given that the model had not been specifically trained to the operating context of a Computer Science department, it was somewhat surprising how relevant many of the general suggestions were. The model struggled, however, with specific suggestions for domain specific issues. The model also made a number of suggestions that were not feasible or were outside the control of an individual instructor, such as changing ECTS credit points, or the duration of a course.

The model struggled with making suggestions when there were contradictory inputs. For example, if there were two student evaluations and one student mentioned they liked online teaching and one did not, the model selects only one part of the feedback, mentioning that *all* students liked online teaching. On a more general note, there is a common pattern; when the input to the model is *sparse* (i.e., not many evaluations) the model generates non-actionable items, or in other words, hallucinates new problems. When the input is too *dense* (i.e., many evaluations), we notice that the model has a harder time prioritising which feedback to generate.

## 4.3. Observations regarding Appropriateness.

Overall the model produced output that was appropriate to share with academics, however there were issues with losing some of sentiment from the student feedback. There were multiple instances of names being included, despite the model being explicitly prompted not to include names. In all such cases the mentions had positive sentiment, so this did not negatively affect the appropriateness of the feedback, but it undermines the trust in the model not to convey negative sentiment outputs in the future.

There were instances in which specific comments with strong sentiment, both positive (e.g., "super helpful") and negative (e.g., "hopeless"), were not incorporated in the synthesised output. A consequence of this was that the resulting output did not convey the strength of that sentiment, instead seeming much like the output for more mild sentiment inputs. While this meets the objective of providing appropriate output, it undermines the factuality and actionability of the outputs by not conveying to the academics how the students are actually responding to their teaching.

## 5. CONCLUSION

In this work, we present a proof-of-concept leveraging open-source generative AI to automatically synthesise actionable teacher feedback from student course evaluations. We prompted an out-of-the-box open-source LLM to summarise course evaluations and give actionable feedback based on this summary. The model generates output consistently based on three common formats. Despite the simplicity of the model, our findings suggest the feasibility of using open-source models to generate summarised actionable feedback from course evaluations as a cost-effective approach to supporting teachers' development. However, we believe further improvements are necessary to improve the effectiveness and accuracy of the generated feedback. With continued refinement, integrating generative AI into educational settings could significantly contribute to improving teaching practices and supporting educators' professional growth.

## 6. ACKNOWLEDGEMENTS

This work was supported by a research grant (VIL57392) from VILLUM FONDEN.
This work was supported by a seed funding grant from the Aalborg University Institute for the Advanced Study of Problem Based Learning.
This work was approved by the Aalborg University Human Research Ethics committee, with approval number 2023-505-00102.
We thank Hans Erik Heje and Karl-Emil Grarup Hertz for their input to the initial experimental implementation, and the anonymous reviewers for the feedback to this manuscript.